\documentclass[10pt,emptycopyrightspace]{ewsn-proc}

\usepackage{graphicx}
\usepackage{balance}
\usepackage{comment}
\usepackage[binary-units=true]{siunitx}
\usepackage{subcaption}
\usepackage{mwe}
\DeclareSIUnit \Kbps {Kbps}
\DeclareSIUnit \Mbps {Mbps}
\newcommand{\textapprox}{\raisebox{0.5ex}{\texttildelow}}

\numberofauthors{2}

\author{
\alignauthor Charles Lockie\\Ioannis Mavromatis\\Aleksandar Stanoev\\Yichao Jin \\
    \affaddr{Bristol Research and Innovation Laboratory (BRIL)}\\
    \affaddr{Toshiba Europe Ltd., Bristol, UK}\\
    \email{Ioannis.Mavromatis@toshiba-bril.com}
    \email{Yichao.Jin@toshiba-bril.com}
\alignauthor George Oikonomou \\
    \affaddr{Electrical and Electronic Engineering}\\
    \affaddr{University of Bristol, Bristol, UK}\\
    \email{G.Oikonomou@bristol.co.uk}
}

\title{Securing Synchronous Flooding
Communications: An Atomic-SDN Implementation}

\begin{document}

\maketitle

\begin{abstract}
Synchronous Flooding (SF) protocols can enhance the wireless connectivity between Internet of Things (IoT) devices. However, existing SF solutions fail to introduce sufficient security measures due to strict time synchronisation requirements, making them vulnerable to malicious actions. Our paper presents a design paradigm for encrypted SF communications. We describe a mechanism for synchronising encryption parameters in a network-wide fashion. Our solution operates with minimal overhead and without compromising communication reliability. Evaluating our paradigm on a real-world, large-scale IoT testbed, we have proven that a communication layer impervious to a range of attacks is established without sacrificing the network performance.
\end{abstract}

%
%

%
\keywords{IoT, Synchronous Flooding, Security, Bluetooth, CCM}

\section{Introduction}

The Internet of Things (IoT) describes the network of interconnected physical objects~\cite{iot_summary}. Wireless connectivity has become a driving force for IoT deployments and their stable operation. Wireless devices, such as sensors, actuators or controllers, can operate independently and communicate with each other and the wider Internet, forming huge distributions of networks~\cite{wirelessSurvey}.

A core technology in wireless networking is flooding~\cite{networkFlooding}. Network flooding is a pattern of communication where a message from one device is re-broadcasted to all other devices until it reaches the furthest device in the network. IoT devices are typically low cost, low power and resource-constrained~\cite{iot_summary_2}. Thus, lightweight communications are crucial in IoT deployments. Synchronous Flooding (SF)~\cite{synchronousFlooding} protocols enable rapid network flooding while solving the collision and scheduling problems from traditional flooding~\cite{networkFlooding}. Facilitated by Concurrent Transmission (CT), multiple synchronised devices simultaneously transmit the same message to their neighbours, allowing for rapid flood propagation across networks.

SF protocols have been proven to improve reliability, reduce latency and outperform traditional multi-hop mesh network protocols~\cite{mavromatis2022reliable}. However, a significant factor limiting the broader adoption of SF protocols is the lack of security features. Addressing these limitations will be the focus of this paper. In this work, we present a design paradigm for encrypted SF communications. The unique characteristics of secure SF will be considered and demonstrated on an existing SF protocol called Atomic~\cite{baddeley2019atomic}.

Secure flooding communication have been considered in the past~\cite{secureFlooding1,secureFlooding2,secureFlooding3}. These works, even though not strictly SF-related, describe how reusing the given sequence numbers (used to synchronise the nodes) can be used to run replay attacks. They also present counter-measurements for preventing them. To the best of our knowledge, no other works have proposed solutions for secure SF protocols. Building on top of the SF mechanisms introduced in Atomic, we intend to introduce the various challenges identified and provide solutions around them.

Security implementations can be found in other wireless protocols in the IoT literature. For example, Bluetooth Mesh~\cite{bleMESH} protects against trash-can attacks using multiple keys and a multi-layer security. Furthermore, the protocol's sequence numbers and an Initialisation Vector (IV) are utilised to protect against replay attacks. Similarly, our approach is based on multipart IVs and encryption keys to ensure protection against these attacks. Bluetooth Mesh, hopping between three channels in a pre-defined sequence makes it vulnerable to jamming attacks~\cite{bleMESH}. Atomic, employing a more diverse channel hopping mechanism, improves resilience to interference and can better protect against jamming attacks.

ZigBee configures the network security from a central coordination node~\cite{zigbeeSecurity}. This node is responsible for generating master and link keys and disseminating them to the rest of the network. Sequence numbers are associated with the key instances and are used to identify them. ZigBee operates on a hop-by-hop security strategy. Each packet received is decrypted, its integrity is verified and later is re-encrypted and sent to the next destination. Such an operation is prune to transport layer attacks (e.g., de-synchronisation attacks) from an internally compromised station~\cite{zigbeeAttack}. Our implementation considers a double-encryption mechanism that ensures only nodes with specific keys can access the payload of a packet.

Using an already-existing solution for Atomic is not applicable. The problem with SF protocols is the tight time synchronisation required between the devices. All current solutions found in Bluetooth Mesh~\cite{bleMESH}, ZigBee~\cite{zigbeeSecurity} and WiFi~\cite{wifiCCMP} introduce increased overhead that leads to network desynchronisation and thus reduced reliability. Our solution will tackle this problem and provide secure communications without compromising the network performance.

The rest of the paper is organised as follows. Sec.~\ref{sec:sfSecurity} describes the SF operations,  its security considerations, and the unique problems SF communications introduce. Our design solutions addressing these problems and our implementation are found in Sec.~\ref{sec:design}. Our large-scale reliability evaluation of the secure SF implementation is found in Sec.~\ref{sec:investigation} and our final remarks are presented in Sec.~\ref{sec:conclusions}.

\section{Synchronous Flooding and Security}\label{sec:sfSecurity}

``Secure'' networking entails \textbf{encrypted} and \textbf{authenticated} messaging~\cite{iotSecurity}. Encrypting data ensures confidentiality - only trusted devices can access the original message. Authenticating data ensures messages cannot be tampered with, allowing receivers to be confident in the origin of a message. As discussed earlier, the lack of security in state-of-the-art SF protocols is very prominent in the literature. This section will briefly describe Atomic's basic operation and the security mechanisms introduced in our implementation.

\subsection{Network Synchronisation and SF}\label{subsec:networkSynchronisation}
All nodes in a CT network broadcast packets simultaneously and on the same carrier frequency. SF, building on this idea, provides a solution to the contention problem and can support one-to-all communication within a single flood, minimising the latency and enhancing reliability.

The tight network synchronisation required is achieved by synchronising the clocks of all nodes. Every time a flood is received, the nodes calculate a ``reference time'', i.e., the exact value of the initiator's clock. A ``Relay Counter (RC)'' is encapsulated in each packet, counting the number of hops within the flood. Knowing the number of hops and that each hop takes a fixed amount of time, nodes can calculate when the flood started and synchronise their clocks for each transmission.

\begin{figure}[t]
    \centering
    \includegraphics[width=1\columnwidth]{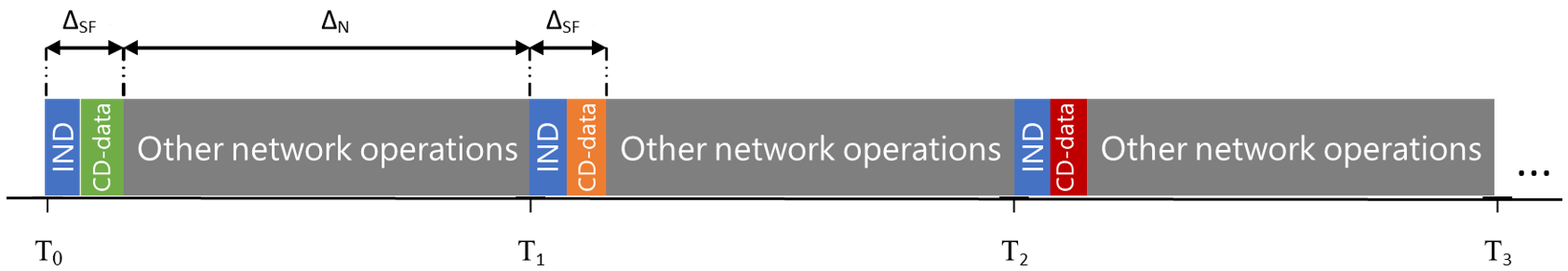}
    \caption{A time-sliced SF control to maximise network resource utilisation. The different data periods represent the different core data patterns.}
    \label{fig:timeslicedSF}
\end{figure}

\subsection{Atomic SF and Data Transfer Modes}\label{subsec:atomicSFData}
Atomic builds on the above-mentioned ideas and provides a low-overhead SF implementation. Multiple floods can be scheduled in pre-defined periods called ``epochs'', decoupled from the rest of the network operation (Fig.~\ref{fig:timeslicedSF}). Also, each flood, described as ``phase'',  has a fixed maximum number of re-transmissions (hops).

Atomic provides three core patterns (Fig.~\ref{fig:timeslicedSF}) for data transfer~\cite{baddeley2019atomic}:
\begin{itemize}
\item Point-to-point (P2P): A flood disseminates information from one node to another, e.g., sending control instructions to individual nodes. Other nodes participate but do not process the packet.
\item Point-to-multipoint (P2MP): The initiator broadcasts a message to all nodes, e.g., used for IND phases or network-wide firmware updates.
\item Multipoint-to-point (MP2P): Nodes flood a message back to a receiver (``uplink''), e.g., sensor data sent to a server from all IoT nodes.
\end{itemize}

\subsection{Network Join and Channel Hopping}\label{subsec:networkJoin}
\begin{figure}[t]
    \centering
    \includegraphics[width=1\columnwidth]{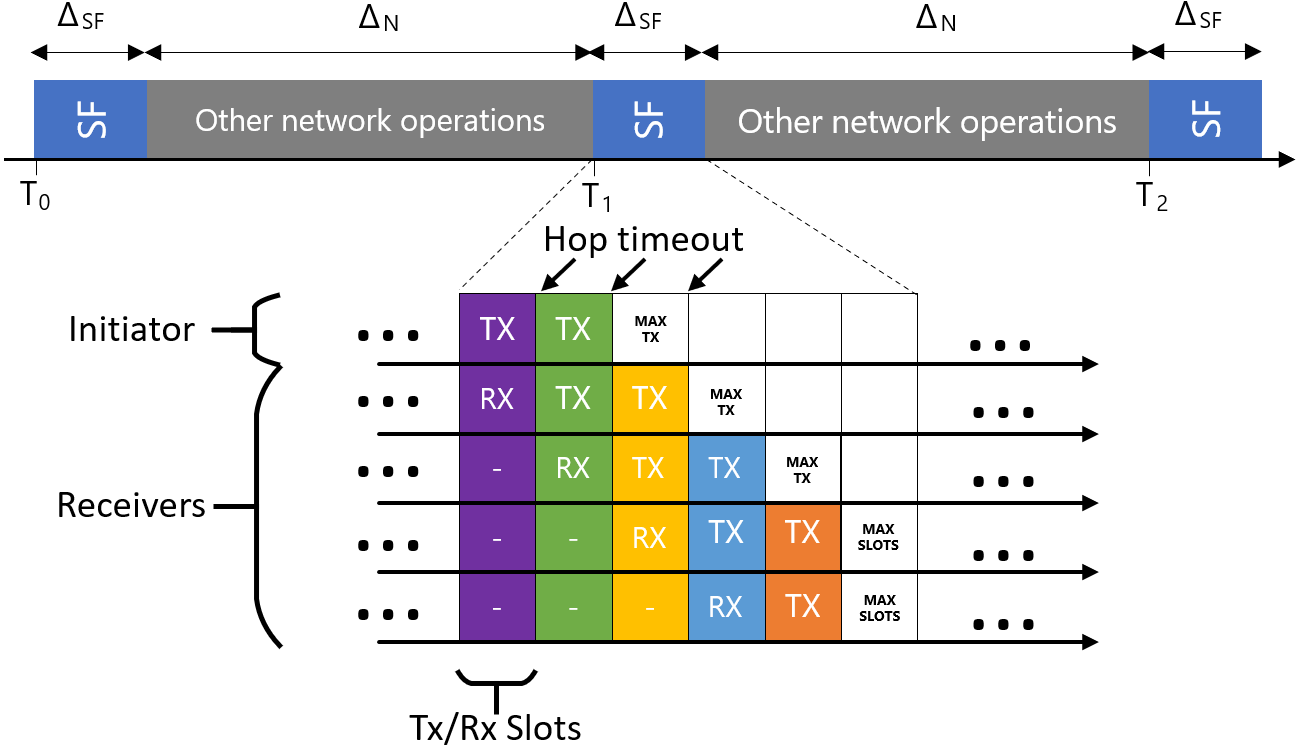}
    \caption{Example of SF used in Atomic. Back-to-back transmissions flood the network with minimal latency.}
    \label{fig:atomicOperation}
\end{figure}

Nodes joining the network require the hopping sequence and the ``reference time'' to know when to expect a flood. The hopping sequence is ``seeded'' by the ``Epoch Counter (EC)'' and controlled by a pseudo-random generator. All Atomic epochs start with a P2MP flood of a single packet that is called Indicator (IND) (Fig.~\ref{fig:timeslicedSF}). IND, amongst other data, contains the current EC. The node initiating the flood is called the ``Initiator'' (Fig.~\ref{fig:atomicOperation}).

The hopping sequence poses a challenge: nodes waiting to join will only receive the IND if they are on the same channel. A workaround is achieved by ensuring some specific channels are ``guaranteed to occur''. Thus, nodes listening to them will be assured to receive the IND, determine the reference time, and successfully join the network. Additionally, as described in Sec.~\ref{subsec:wirelessSecuritySF}, joining a secure network requires knowledge of an IV and a key to decrypt the IND packet. In Sec.~\ref{subsec:joinSecure} we describe our design considerations for that.

\subsection{Wireless Security for SF}\label{subsec:wirelessSecuritySF}
Two key aims of security are Authenticity and Confidentiality~\cite{wireless_security_survey}. Authenticity aims to verify the sender's identity, and that the message contents have not been modified. Confidentiality ensures a message can be read or copied only by the sender and the recipient. Encryption is the process of converting message data, i.e., a ``plaintext'', into a cipher, i.e., ``ciphertext'', guaranteeing confidentiality.

Data encryption is usually achieved by two methods, i.e., with block ciphers or stream ciphers. Both are symmetric-key ciphers. A block cipher breaks down plaintext messages into fixed-size blocks before converting them into ciphertext using a key, while a stream cipher breaks a plaintext message down into single bits, which then are converted individually into ciphertext using key bits~\cite{securityGeneric}.

Our approach is based on the widely-adopted  Counter with Cipher Block Chaining-Message Authentication Code (CCM) mode~\cite{ccm_rfc}, used in a number of protocols such as Bluetooth Mesh~\cite{bleMESH}, ZigBee~\cite{zigbeeSecurity}, and WiFi~\cite{wifiCCMP} (with minor variations for each). Its implementation allows both authenticity and confidentiality to be achieved with a single algorithm.

The CCM encryption is divided into two individual parts: the Counter Mode (CTR) and the Cipher Block Chaining-Message Authentication Code (CBC-MAC). The CTR is used to encrypt the plaintext to ciphertext using a key and an incremental counter, and CBC-MAC generates a \SI{4}{\byte} long Message Authentication Code (MAC) using a constant IV of zeros and a key. The MAC is appended at the end of the packet allowing receivers to verify the integrity of a received message. MAC is otherwise known as Message Integrity Check (MIC) for Bluetooth communications. Finally, the cipher keys used in CCM are based on the Advanced Encryption Standard (AES)~\cite{aes}. More in-depth information about the operation of CCM-AES can be found in~\cite{ccm_rfc,securityGeneric}.

\subsection{The Challenge of Time Drifting in SF}\label{subsec:drift}
Atomic, similarly to other SF protocols, requires all nodes to transmit at the same time. Any time difference leads to overlapping transmissions and thus destructive interference on the channel. Oscillator drifts observed in embedded devices (due to environmental factors and ageing) are compensated in Atomic by synchronising a receiving node every time it receives a hop in each flood. As time passes after synchronisation, the drift increases. Any flood retransmissions must be performed while drift is below \SI{0.5}{\micro\second}. This limits the number of times a node can transmit reliably. If the time between hops increases, the reliability decreases, leading to increased Packet Error Rate (PER).

Moreover, AES computation takes the most signification proportion of time required for CCM. As measured~\cite{softwareCCM}, the computation time on hardware is \SI{80}{\micro\second} and on software is almost 20-fold larger (\SI{1556}{\micro\second}).  For Atomic increasing the executing time increases the drift and thus reduces the reliability.

\subsection{IVs and Nonces}\label{subsec:ivsNonces}
IVs are used as input to the AES block when generating a keystream. The CCM RFC~\cite{ccm_rfc} specifies that the IV must be a number-once (nonce). That implies that one IV is used to encrypt a single plaintext. Reusing an IV results in encrypting many messages with the same keystream, which is against the general rule of cryptography, i.e., a keystream should never be used more than once. It can also lead to ciphertext-only attacks, e.g., a many-time-pad attack.

Introducing a layer of security on top of Atomic (CCM) adds extra overhead in the communication channels. As described in Sec.~\ref{subsec:drift}, increased drift for larger floods reduces the reliability of Atomic. Other wireless protocols using CCM (e.g., WiFi) include the IV in the header sent with each packet and utilise a pre-shared key for the decryption. The decryption starts once the packet is received. This implies that for a AES-128 (\SI{128}{\bit}) block cipher, an additional \SI{16}{\byte} are added to all frame headers. Moreover, sending the IV in-packet and decrypting after reception increases the overall time required even further. These issues will all manifest in more desynchronisations. If synchronisation between all Atomic nodes fails, the PER will rise to $100\%$ (total failure). Therefore, workarounds for the above problems must be considered when securing Atomic.


\section{Design and Implementation}\label{sec:design}

As described in Sec.~\ref{sec:sfSecurity}, to maintain the benefits of SF and Atomic, we should minimise any additional overhead while maintaining the tight synchronisation between the nodes. Furthermore, we need to ensure the authenticity and confidentiality. In this section we describe our design decisions that can accommodate the above.

\subsection{CCM On-the-fly}\label{subsec:ccmOnTheFly}
As discussed in Sec.~\ref{subsec:drift}, the computation of CCM on software is significantly longer than on hardware. Therefore, a hardware peripheral is mandatory. Based on that, our wireless interface of choice was the Nordic nRF52840~\cite{nRF52840}. nRF52840 provided a hardware CCM peripheral that can asynchronously generate the keystream prior to transmission and XORs the bits as they are sent. Similarly for the reception, the keystream is generated by the CCM peripheral and the ciphertext is XORed generating the plaintext message. This allows ``on-the-fly'' encryption and decryption.

The CCM peripheral for Nordic nRF52840 was designed for Bluetooth Low Energy (BLE)~\cite{nRF52840}. As a result, it expects a specific packet structure different from Atomic~\cite{baddeley2019atomic}. The main difference was the lack of a \SI{1}{\byte} control field for the frame size. Atomic, operating with fixed payload lengths, did not require this field in its header. To accommodate that, Atomic was extended to encapsulate the size in the frame header.

\subsection{Synchronising the IVs}

\begin{figure}[t]
    \centering
    \includegraphics[width=1.0\columnwidth]{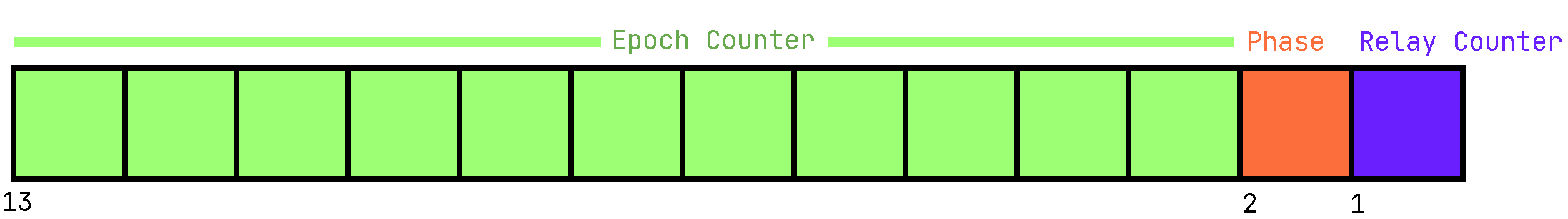}
    \caption{The chosen IV structure for enabling network synchronisation. Each block is \SI{1}{\byte} long.}
    \label{fig:iv_structure}
\end{figure}

CTR mode relies on a pre-generated keystream to encrypt and decrypt packets. As discussed in Sec.~\ref{subsec:ivsNonces}, sending the IV on a per-packet basis (like in WiFi) is not an option for Atomic. Therefore, both transmitters and receivers must agree on the IV ahead of time. We solve this problem by creating a multipart IV based on the Atomic EC, RC and Phase Counter (PC), as shown in Fig.~\ref{fig:iv_structure}. As illustrated in Sec.~\ref{subsec:networkJoin}, all nodes synchronise the EC value in the IND phase. Moreover, when a node participates in a flood as a transmitter, it rebroadcasts the same packet multiple times (Fig.~\ref{fig:atomicOperation}). During this operation, a transmitter updates its RC value, reflecting the number of re-transmissions. At the same time, the receivers keep track of the expected RC, this being the index for the channel hopping sequence (Sec.~\ref{subsec:networkSynchronisation}).

Combining EC and RC provides a unique, synchronised IV. This IV can be later used as the key and the incremental counter required for CTR. Furthermore, EC can be preserved across hardware restarts (e.g. by storing it in a non-volatile flash), providing a globally unique IV until the value overflows. A 64-bit EC is used in our implementation, allowing for $2^{64}$ epochs per key. Finally, in the case multiple Atomic phases occur per epoch, the PC is additionally included in the IV (Fig.~\ref{fig:iv_structure}).

The requirement for CCM encryption is that both the transmitter and the receiver use the same IV. This is acceptable for Atomic P2P and P2MP data patterns since a single message is being flooded across the entire network. In the case of MP2P, this poses a problem. More than one node may try and initiate a flood, each with different payloads. This entails encrypting different payloads with the same IV, constituting a many-time-pad attack. Our design breaks confidentiality on all plaintexts encrypted using that IV. Authenticity is still valid, as CBC-MAC does not rely on the pre-agreed IV.

\subsection{Double Encryption through Device Keys}

The above behaviour may be acceptable for many use cases. However, if confidentiality is required for MP2P, nodes must encrypt their payloads separately. This could be achieved by a ``device key'' known only to specific nodes. Encrypting the payload before transmission allows packets to be flooded but not accessed by the entire network. For example, a vendor-specific firmware update could be scheduled in such a way. All nodes flood the packets received, but only a subset with the correct keys can decrypt them.

This secondary encryption is handled by a layer higher than the Atomic MAC (e.g., the application). What is more, additional time should be provisioned for the secondary encryption. As long as encryption and decryption occur outside the SF Atomic slot, SF operations are unaffected.

Key provisioning is an active area of research~\cite{bleKeysResearch}. The focus of this paper is outside of this scope. For our implementation, we assigned a single shared key to all devices before our experimentation. For a real-world implementation, a key provision and distribution mechanism like the one found in BLE can be used to distribute new secret keys to all devices.

\subsection{Joining a Secure Network}\label{subsec:joinSecure}
Joining the Atomic network requires the current EC (Sec.~\ref{subsec:networkJoin}). Encrypting the IND flood will block nodes from receiving the current EC and calculating the IV. To overcome that, a known constant IV must be used, thus breaking confidentiality. For our implementation, a constant value of 0 was chosen. This still achieved authenticity in Atomic. The contents of the IND packet do not reveal any information that risks security if known.

For real-world deployment, a solution for the above limitation can be an out-of-band authentication mechanism responsible for exchanging the EC information. This will enable a multi-factor authentication scheme with secondary verification mechanisms operating through a separate communication channel. An example of such a solution can be found in~\cite{outOfBandAuthentication}, where a blockchain-enabled mechanism is proposed.

\subsection{Implementation Considerations}\label{subsec:implementation}
The CCM peripheral on the nRF52840 SoC can generate the keystreams in the same amount of time taken for the radio ramp-up ($<\SI{50}{\micro\second}$ as described in the specification~\cite{nRF52840}). The encryption begins as soon as the keystream generation is completed. The transmission occurs as soon as the radio ramps up, coinciding with the encryption's commencement. Similarly, the decryption begins once the packet's payload portion is received. The radio peripheral generates an event once the address field has
been received initiating the decryption phase.

The peripheral is capable of supporting four bitrates, i.e., \SI{125}{\Kbps}, \SI{500}{\Kbps}, \SI{1}{\Mbps}, and \SI{2}{\Mbps}. These are the four bitrates supported by Atomic, so the peripheral can support all current PHYs in our implementation. The  CCM peripheral uses the same clock source to operate synchronously with the radio at any bitrate. If enabled simultaneously, the CCM operation completes concurrently with the radio, thus allowing us to operate in the tightly time-synchronised environment required for SF.

With regards to the packet structure, as described in Sec.~\ref{subsec:ccmOnTheFly}, \SI{1}{\byte} is in the header, describing the frame size and replicating the BLE frame structure in Atomic. In addition, the CCM peripheral introduces another \SI{4}{\byte} of a payload extension (MIC) appended at the end of the payload and before the Cyclic Redundancy Check (CRC). While testing our implementation, a behaviour was observed that when the frame size field is corrupted, the node is desynchronised. If the corrupted value is significantly higher than the intended value, the receiver can overrun the end of the hop, causing scheduling to fail. We solved that by configuring a ``maximum packet length'' set in the radio before the reception. This limits the receive time by terminating the reception after a fixed number of bytes and discards failed packets. Finally, Atomic can schedule phases knowing the exact time required for packet transmission. These timings are experimentally calculated as the fixed time per payload bit, plus a constant time for the preamble, header and CRC. These constants were measured for each PHY for the new packet structure.

\section{Experimental Investigation}\label{sec:investigation}

In this section, we will analyse the performance of Atomic in two scenarios: unencrypted and encrypted traffic. Compared to other technologies, the benefits of Atomic have already been discussed in our prior work~\cite{mavromatis2022reliable}. Our experimentation is conducted on the UMBRELLA testbed.

\begin{figure}[t]
    \centering
    \includegraphics[width=1\columnwidth]{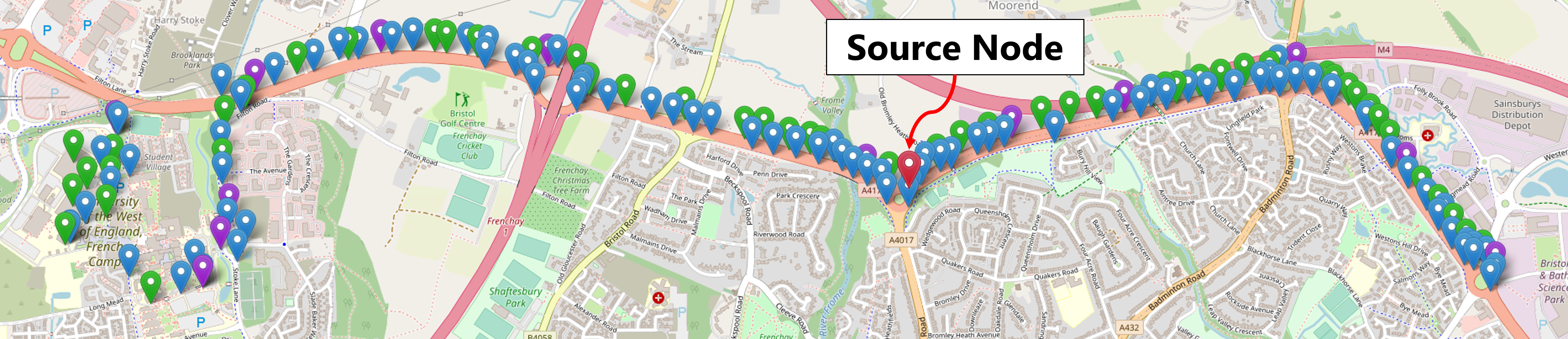}
    \caption{The UMBRELLA network. All nodes are installed on public lampposts across a road of \textapprox7.2km. The red node is our experiment source node. The rest of the nodes are all equipped with an Nordic nRF52840.}
    \label{fig:umbrella_network}
\end{figure}

The UMBRELLA testbed is installed across a \textapprox7.2km stretch of road (Fig.~\ref{fig:umbrella_network}).  UMBRELLA nodes are equipped with nRF52840 interfaces. Between the interface and the dipole antenna exists a Skyworks RF Front-End Module, integrating a Low Noise Amplifier (LNA) and Power Amplifier (PA). This results in \SI{22}{\dB} of TX power gain, and increases RX sensitivity up to \SI{6}{\dB}.

At the beginning of an experiment, all nodes are flashed with Atomic's firmware simultaneously. Atomic was configured in P2MP mode, i.e. the source node shown in Fig.~\ref{fig:umbrella_network} (initiator), floods fixed-size packets across all nodes. Recipients count the number of successful packets received, i.e., successful CRC and MIC, while the source node logs the total number of packets sent. At the end of an experiment, the experiment logs are collected on our server, where the PER is calculated. All tests were repeated for $10$ times and \SI{3000}{\second} each. Atomic is configured to schedule epochs at \SI{500}{\milli\second} intervals, resulting in $6000$ floods per experiment. Finally all four available PHYs were evaluated, i.e., \SI{125}{\Kbps}, \SI{500}{\Kbps}, \SI{1}{\Mbps}, and \SI{2}{\Mbps} and four different payloads, i.e., \SI{20}{\byte}, \SI{50}{\byte}, \SI{100}{\byte} and \SI{200}{\byte}.

\subsection{Results and Discussion}
\begin{figure*}[t]
\centering
\begin{subfigure}[t]{1\columnwidth}
    \centering
    \includegraphics[width=\columnwidth]{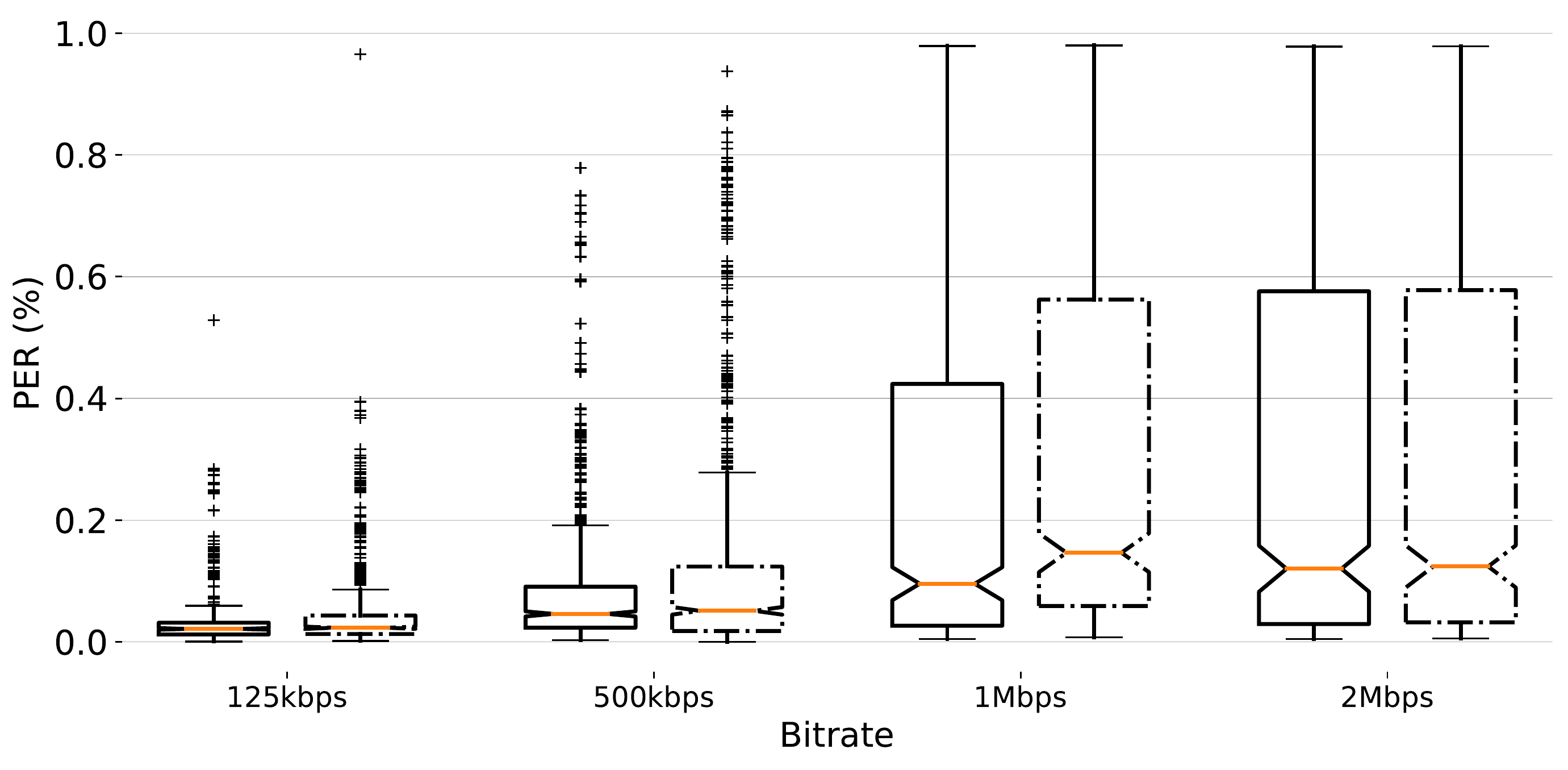}
    \caption{PER distribution for a \SI{20}{\byte} payload.}
    \label{subfig:payload20}
\end{subfigure}
~ 
\begin{subfigure}[t]{1\columnwidth}
    \centering
    \includegraphics[width=\columnwidth]{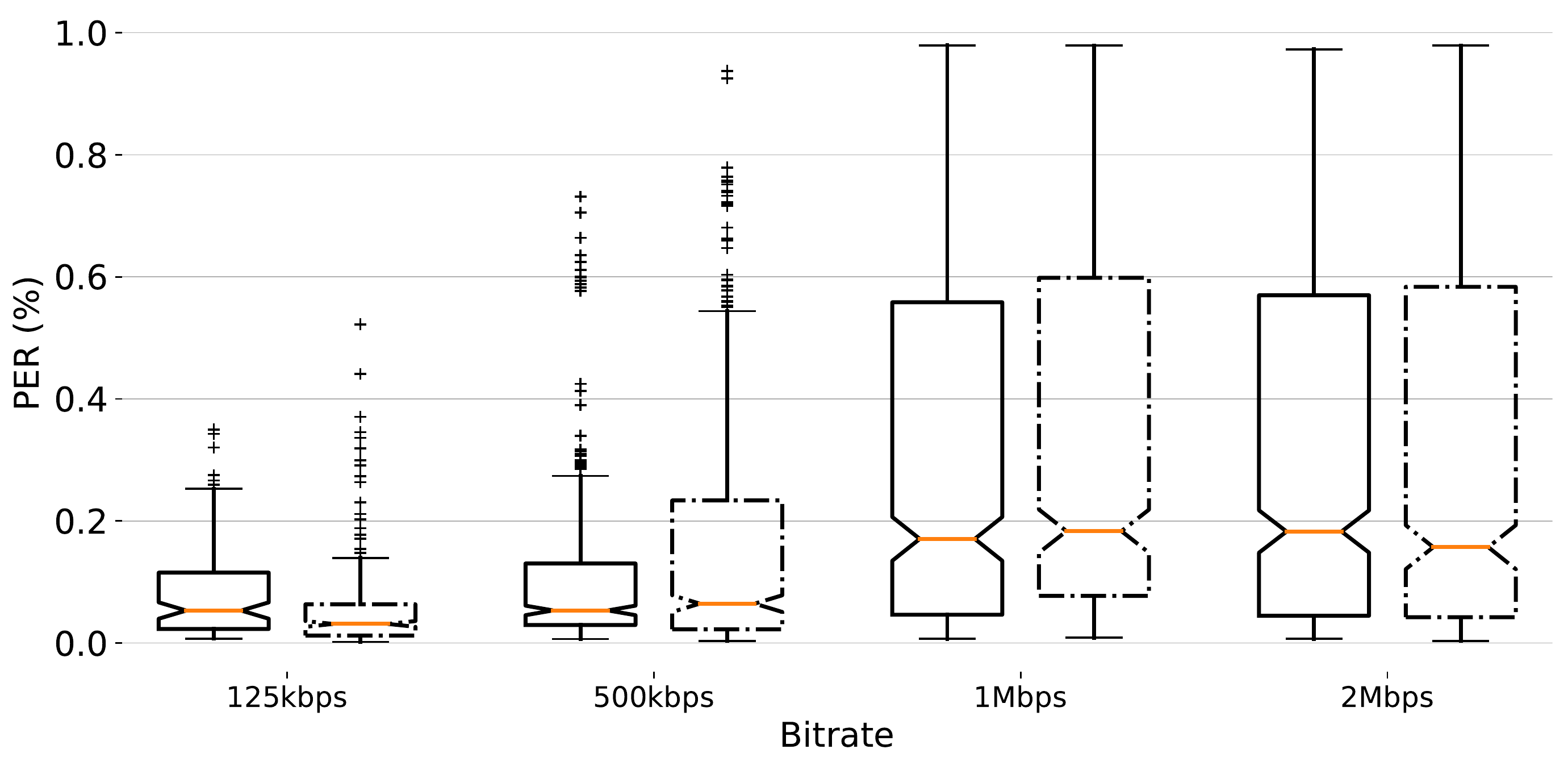}
    \caption{PER distribution for a \SI{50}{\byte} payload.}
    \label{subfig:payload50}
\end{subfigure}

\begin{subfigure}[t]{1\columnwidth}
    \centering
    \includegraphics[width=\columnwidth]{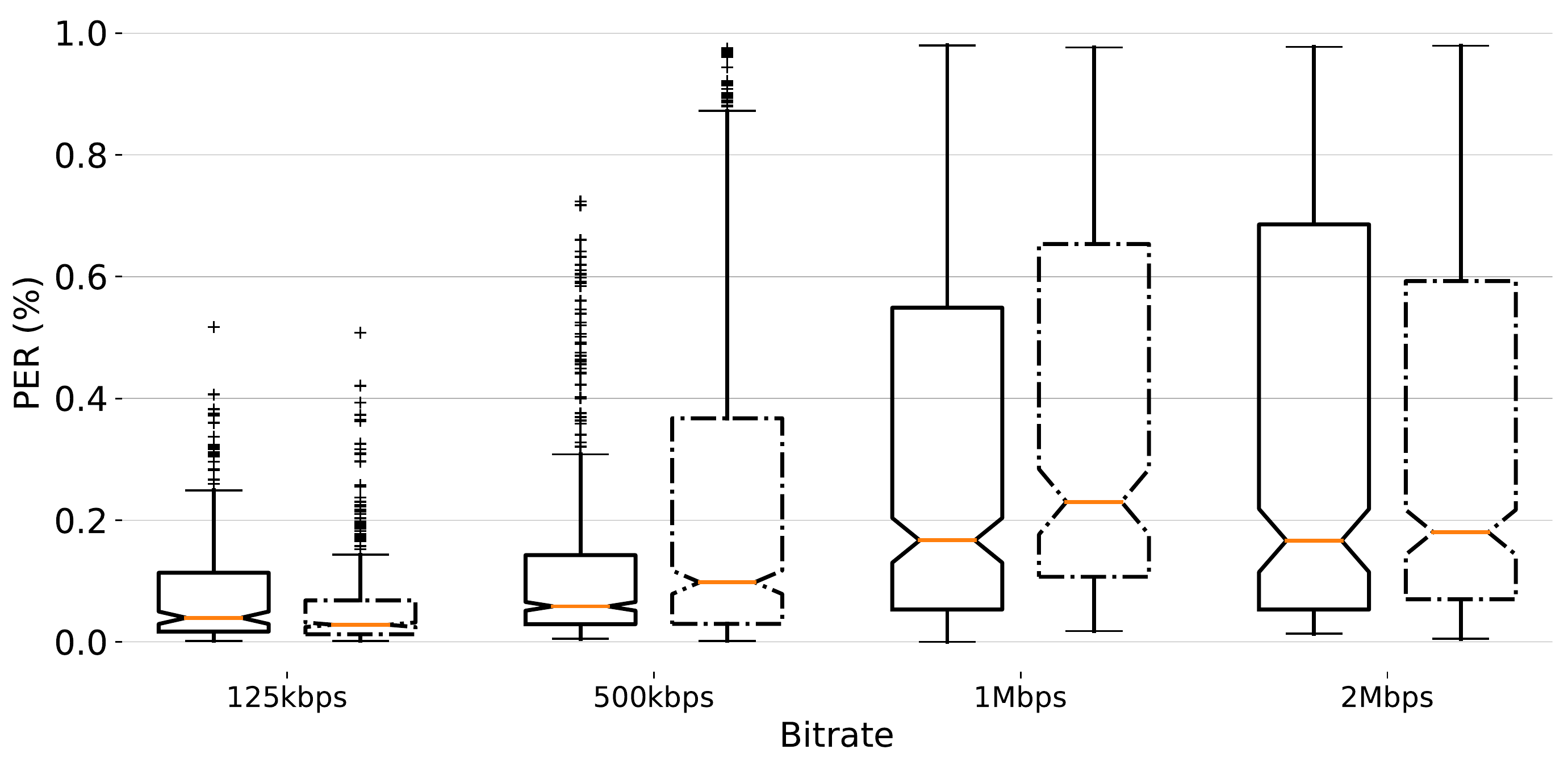}
    \caption{PER distribution for a \SI{100}{\byte} payload.}
    \label{subfig:payload100}
\end{subfigure}
~ 
\begin{subfigure}[t]{1\columnwidth}
    \centering
    \includegraphics[width=\columnwidth]{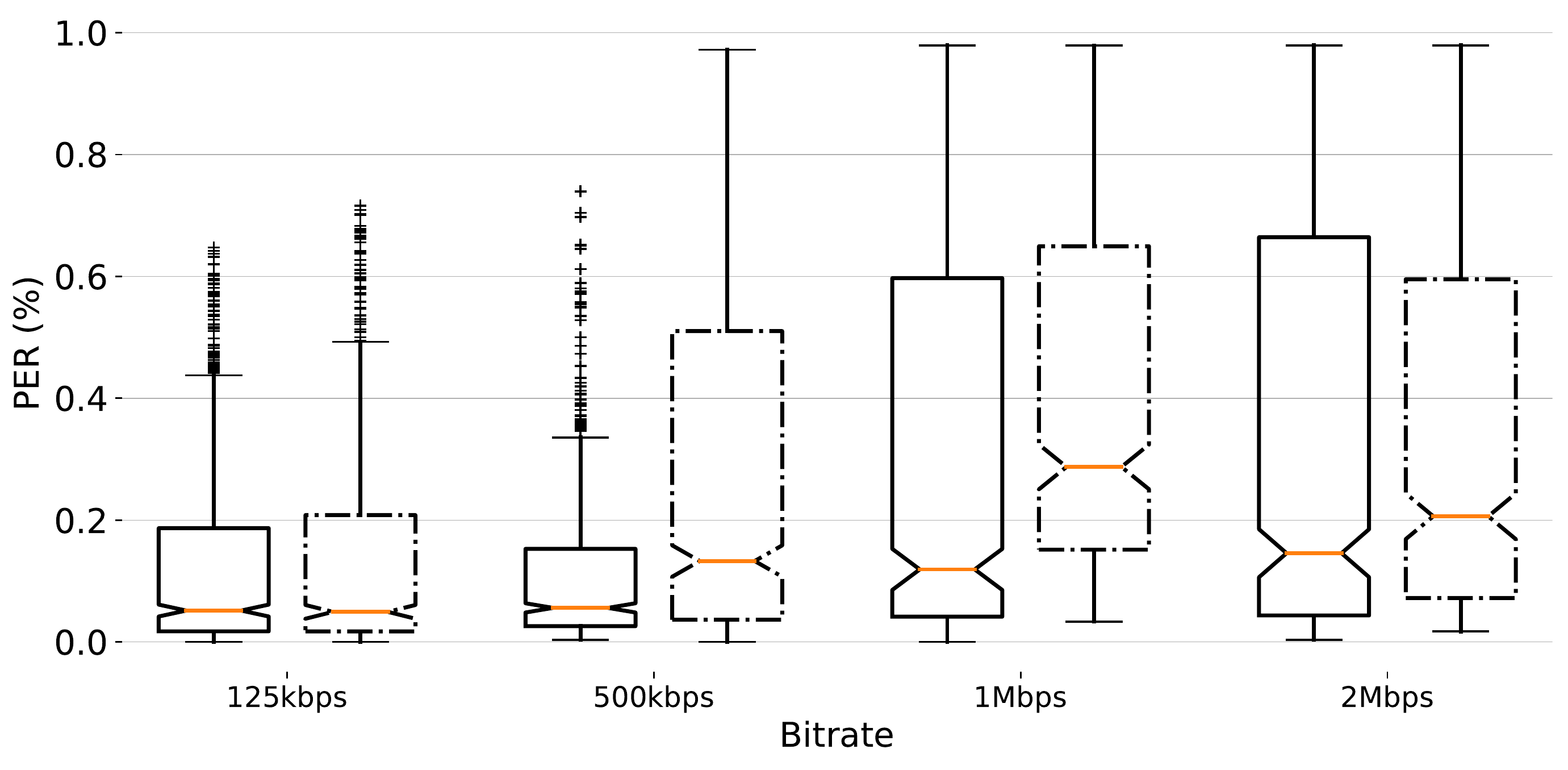}
    \caption{PER distribution for a \SI{200}{\byte} payload.}
    \label{subfig:payload200}
\end{subfigure}
\caption{PER performance distribution unecrypted and encrypted Atomic, and different payload sizes and PHYs. The dotted boxplots (right) present the encrypted Atomic. The solid line boxplots (left) present the unencrypted experiments.}\label{fig:per}
\end{figure*}

Fig.~\ref{fig:per} summarises our performance investigation. We compare the perceived PER for all available PHYs, four payloads and two types of Atomic traffic, i.e., unencrypted and encrypted. Overall, we see that higher bitrates increase the PER. This is expected as lower PHYs introduce forward-error-correction mechanisms and more symbols-per-bit. What is more, a higher variance for higher bitrates is also expected. Increased bitrate implies a shorter transmission range and affects different parts of the network disproportionately: nodes in dense clusters can reach many neighbours, whereas nodes at the periphery do not. Sparse nodes experience higher PER, and thus the increased variance. Moreover, UMBRELLA being an urban testbed, is affected by external interference. For higher bitrates, this effect is more prominent, introducing increased PER.

Considering the unencrypted and encrypted traffic, a general observation is that encrypted Atomic performs slightly worse, but overall PER distributions are broadly comparable for all PHYs. Apart from the \SI{500}{\Kbps} all other results are within a \textapprox$5\%$ margin. The five extra bytes added to the frame lead to more corrupted bits, increasing the PER. When considering the node locations, it was observed that nodes close to the source present almost identical performance. In contrast, nodes several hops away endure an increased PER. The increased number of hops introduces slight desynchronisation and leads to increased PER combined with the increased overhead. Moreover, comparing the PER for the different payloads, we can see that as the payload size increases, the variance of the PER increases as well. This is expected behaviour. As more bits are sent per transmission, the noisy UMBRELLA channels lead to more collisions and, thus, more corrupted packets.

As seen, the PER difference between unencrypted and encrypted traffic and the \SI{500}{\Kbps} PHY increases significantly compared to the other PHYs. An explanation for that is the timing changes discussed in Sec.~\ref{subsec:implementation}. If these timings are marginally off, Atomic will wrongly estimate the time taken for a TX. Our results show a PER increase of $10\%$, suggesting a small timing error. Nodes at higher hop counts will see an increased PER or total failure, whereas nodes closer to the initiator (capable of receiving a flood in a small number of hops) will see no reduction. We intend to recalculate these timing constants for \SI{500}{\Kbps} PHY and fix this desynchronisation issue.

\subsection{Benefits of Security}

Secure Atomic can mitigate against several attacks. Firstly, packet interception attacks involve listening to the communication channel, receiving any messages sent, and decoding them. Secure Atomic prevents that, as parties can only decipher encrypted packets with the encryption key. Packet injection attacks involve a malicious party sending packets as if they are a participant in the flood. This is not possible anymore, as nodes will only be able to receive messages encrypted with the secure key, and the rest are discarded. Even in the case of IND floods that any party can decode due to the IV reuse (Sec.~\ref{subsec:joinSecure}), messages must contain a valid MIC to be received, which can only be generated if the key is known.

Packet replay attacks are also prevented. The epoch and hop counters are synchronised network-wide and increment after every reception. Attackers cannot modify these counters, as doing so will invalidate the MIC. Additionally, our design specifies that an IV associated with a key can never be repeated and persists across hardware restarts. This prevents packet replay attacks even after an initiator is restarted. Overall, the secure Atomic implementation not only ensures the prevention of the above attacks but, as our results showed, it performs similarly to the unencrypted version, thus maintaining the benefits of SF communications.

\section{Conclusions}\label{sec:conclusions}
In this paper, we presented our solution for securing Synchronous Flooding communications. More specifically, we described a mechanism for synchronising encryption parameters across an SF network. This allows devices to prepare for encrypting and decrypting SF floods. Our solution accommodates all the different traffic patterns found in a real-world scenario. We demonstrated our solution on a large-scale real-world testbed. As seen, our mechanism achieves secure SF without compromising its advantages, i.e., speed and reliability. We observed that secure communications are achieved, with almost identical performance to the unencrypted implementation. This concludes that secure SF communications is a viable solution for future IoT deployments.

%
%
\section{Acknowledgments}
This work is funded in part by Toshiba Europe Ltd. UMBRELLA project is funded in conjunction with South Gloucestershire Council by the West of England Local Enterprise Partnership through the Local Growth Fund, administered by the West of England Combined Authority.

%

\balance
\bibliographystyle{abbrv}
\bibliography{ewsn-workshops}  

\begin{thebibliography}{10}

\bibitem{baddeley2019atomic}
M.~Baddeley, U.~Raza, et~al.
\newblock {Atomic-SDN: Is synchronous flooding the solution to software-defined
  networking in IoT?}
\newblock {\em IEEE Access}, 7:96019--96034, 2019.

\bibitem{bleKeysResearch}
M.~Cäsar, T.~Pawelke, et~al.
\newblock {A survey on Bluetooth Low Energy security and privacy}.
\newblock {\em Computer Networks}, 205:108712, 2022.

\bibitem{aes}
J.~Daemen and V.~Rijmen.
\newblock {\em The design of {Rijndael}: {AES} --- the {Advanced Encryption
  Standard}}.
\newblock Spring{\-}er-Ver{\-}lag, 2002.

\bibitem{wirelessSurvey}
J.~Ding, M.~Nemati, C.~Ranaweera, and J.~Choi.
\newblock {IoT Connectivity Technologies and Applications: A Survey}.
\newblock {\em IEEE Access}, 8:67646--67673, 2020.

\bibitem{synchronousFlooding}
F.~Ferrari, M.~Zimmerling, et~al.
\newblock {Efficient network flooding and time synchronization with Glossy}.
\newblock In {\em Proc. of ACM/IEEE IPSN 2011}, pages 73--84, 2011.

\bibitem{iot_summary_2}
J.~Gubbi, R.~Buyya, et~al.
\newblock {Internet of Things (IoT): A vision, Architectural Elements, and
  Future Directions}.
\newblock {\em Future generation computer systems}, 29(7):1645--1660, 2013.

\bibitem{networkFlooding}
A.~Hassanzadeh, R.~Stoleru, and J.~Chen.
\newblock {Efficient Flooding in Wireless Sensor Networks Secured with
  Neighborhood Keys}.
\newblock In {\em Proc. of IEEE WiMob 2011}, pages 119--126, 2011.

\bibitem{wifiCCMP}
S.~A. Hoseini, B.~Khodabandeloo, et~al.
\newblock {High Throughput Low Power CCMP Architecture for Very High Speed
  Wireless LANs}.
\newblock In {\em Proc. of Int Symp. on CADS}, pages 59--65, 2010.

\bibitem{secureFlooding1}
D.-J. Huang, K.-J. You, and W.-C. Teng.
\newblock {Secured Flooding Time Synchronization Protocol}.
\newblock In {\em Proc. of IEEE MASS 2011}, pages 620--625, 2011.

\bibitem{bleMESH}
E.~Kalinin, D.~Belyakov, et~al.
\newblock {IoT Security Mechanisms in the Example of BLE}.
\newblock {\em Computers}, 10(12), 2021.

\bibitem{zigbeeSecurity}
S.~Khanji, F.~Iqbal, and P.~Hung.
\newblock {ZigBee Security Vulnerabilities: Exploration and Evaluating}.
\newblock In {\em Proc. of Int. Conf. on ICICS 2019}, pages 52--57, 2019.

\bibitem{zigbeeAttack}
M.~T. Kurniawan and S.~Yazid.
\newblock Mitigation strategy of sinkhole attack in wireless sensor network.
\newblock In {\em Proc. of Int. Conf. on IWBIS 2017}, pages 119--125, 2017.

\bibitem{securityGeneric}
G.~Lackner.
\newblock {A Comparison of Security in Wireless Network Standards with a Focus
  on Bluetooth, WiFi and WiMAX}.
\newblock {\em Int. J. Netw. Secur.}, 15:420--436, 2013.

\bibitem{iot_summary}
I.~Lee and K.~Lee.
\newblock {The Internet of Things (IoT): Applications, Investments, and
  Challenges for Enterprises}.
\newblock {\em Business Horizons}, 58(4):431--440, 2015.

\bibitem{softwareCCM}
X.~Luo, Y.~Qi, Y.~Wan, and Q.~Wang.
\newblock {Overhead Model of CCM for Industrial Wireless Network}.
\newblock In {\em Proc. of IEEE HPCC 2013}, pages 1203--1208, 2013.

\bibitem{iotSecurity}
S.~Marksteiner, V.~J. Exposito~Jimenez, et~al.
\newblock {An Overview of Wireless IoT Protocol Security in the Smart Home
  Domain}.
\newblock In {\em Proc. of IEEE CTTE 2017}, pages 1--8, 2017.

\bibitem{mavromatis2022reliable}
I.~Mavromatis, A.~Stanoev, et~al.
\newblock {Reliable IoT Firmware Updates: A Large-scale Mesh Network
  Performance Investigation}.
\newblock In {\em Proc. of IEEE WCNC 2022}, pages 108--113, May 2022.

\bibitem{secureFlooding2}
T.~Qiu, X.~Liu, et~al.
\newblock {A Secure Time Synchronization Protocol Against Fake Timestamps for
  Large-Scale Internet of Things}.
\newblock {\em IEEE Internet of Things Journal}, 4(6):1879--1889, 2017.

\bibitem{secureFlooding3}
T.~Roosta, W.-C. Liao, et~al.
\newblock {Testbed Implementation of a Secure Flooding Time Synchronization
  Protocol}.
\newblock In {\em Proc. of IEEE WCNC 2008}, pages 3157--3162, 2008.

\bibitem{nRF52840}
N.~Semiconductors.
\newblock {nRF52840 Product Specification, v1.1.}, feb 2019.

\bibitem{ccm_rfc}
D.~Whiting, R.~Housley, and N.~Ferguson.
\newblock {Counter with CBC-MAC (CCM)}.
\newblock RFC 3610, RFC Editor, Sept. 2003.

\bibitem{outOfBandAuthentication}
L.~Wu, X.~Du, et~al.
\newblock {An Out-of-band Authentication Scheme for Internet of Things Using
  Blockchain Technology}.
\newblock In {\em Proc. of Int. Conf. ICNC 2018}, pages 769--773, 2018.

\bibitem{wireless_security_survey}
Y.~Zou, J.~Zhu, et~al.
\newblock {A Survey on Wireless Security: Technical Challenges, Recent
  Advances, and Future Trends}.
\newblock {\em Proc. of the IEEE}, 104(9):1727--1765, 2016.

\end{thebibliography}
\end{document}